\documentclass[a4paper,11pt]{article}
\usepackage{pos}

\title{Multimuon events from cosmic rays with the ALICE detector}

\author*[a]{Mario Sitta}
\onbehalf{(on behalf of the ALICE Collaboration)}

\affiliation[a]{Università del Piemonte Orientale and INFN Torino,\\
  Viale T.Michel 11, Alessandria, Italy}

\emailAdd{mario.sitta@uniupo.it}

\abstract{ALICE experiment at the CERN's Large Hadron Collider, carried out a
cosmic data-taking campaign in the period 2015--2018 (LHC Run 2), for a total of
62.5 days of real observation time. In this contribution, the analysis of these
data is limited to multimuon events, i.e.\ events with more than four detected
muons. In particular, the muon multiplicity distribution (MMD) is studied in the
low-intermediate multiplicity range ($4 < N_{\mu} < 50$), while for
multiplicities $N_{\mu} > 100$, dominated by large statistical fluctuations of
the data, the rate of the high muon multiplicity events (HMM) is measured. The
results are compared with Monte Carlo simulations using three of the main
hadronic interaction models describing the air shower development in the
atmosphere: QGSJET-II-04, EPOS-LHC, and SIBYLL 2.3d, and assuming two extreme
compositions of primary cosmic rays, pure proton and pure iron, representing
respectively the lightest and the heaviest composition. Although the models
have difficulty precisely describing the trend of primary p and Fe as a
function of cosmic ray primary energy, QGSJET-II-04 is the only model that
reproduces reasonably well the MMD and the rate of HMM events assuming a heavy
composition for the entire energy range studied.}

\FullConference{International Symposium on Very High Energy Cosmic Ray Interactions (ISVHECRI2026)\\
5-10 July, 2026\\
Lecce, Italy\\}

\begin{document}
\maketitle

\section{Introduction}
The use of collider detectors to study the atmospheric muons was pioneered by
the LEP experiments ALEPH, DELPHI and L3. All results reported by the LEP
experiments were consistent with the standard hadronic interaction models,
except the observation of high-multiplicity muon-bundle events: even under the
assumption of the highest measured flux and a pure-iron spectrum, the Monte
Carlo models of that time failed to describe the rate of those
high-multiplicity events \cite{Ref:ALEPHRes,Ref:DELPHIRes}.

ALICE (A Large Ion Collider Experiment) \cite{Ref:AliceJINST} is a
general-purpose detector, one of the four main experiments at the CERN LHC. It
focuses on the study of the properties of the Quark--Gluon Plasma (QGP) created
under extreme energy densities in high energy nucleus--nucleus collisions.
Besides the study of heavy-ion physics, since 2009 ALICE developed a cosmic-ray
physics program, which exploits the excellent tracking capabilities of the
Time Projection Chamber (TPC) to detect and reconstruct the muons produced by
the cosmic radiation interacting with the atmosphere.

ALICE is located at Point 2 of the LHC tunnel, 52~m underground and with 28~m
of overburden rock. This depth of rock completely absorbs the electromagnetic
and hadronic components of the cosmic-ray induced air shower, and poses an
energy threshold of 16~GeV for vertical muons \cite{Ref:EnerThr}. The ALICE
solenoid magnet (maximum strength 0.5~T) houses the central-barrel detectors,
where different techniques are exploited to detect and reconstruct all
particles coming out from the primary interaction vertex. A forward muon arm
is located on one side outside the solenoid magnet. A detailed description of
the ALICE detector is given in Ref.\ \cite{Ref:AliceJINST}.

The ALICE Time Projection Chamber (TPC) is the largest detector of this type,
and is the main ALICE tracking device. It has an inner radius of 80~cm, an
outer radius of 280~cm and a total length of 500~cm along the beam axis. In
Run 2, it was read out by multi-wire proportional chambers at both end caps. The
total area to detect cosmic muons is about 25~m$^2$\/; after applying a minimum
length requirement to the reconstructed muon tracks the effective area reduces
to about 17~m$^2$\/.

A dedicated trigger for the cosmic-ray data taking was provided by the Time Of
Flight (TOF), a detector consisting of a cylindrical array of 1638 multi-gap
resistive plate chamber pads arranged in 18 sectors, surrounding the TPC. The
trigger configuration used for the present analysis requires a coincidence
between a signal in a sector of the upper part and a signal in the opposite
lower sector with respect to the central axis of the detector, or in one of the
three adjacent sectors on either side, forming a back-to-back $\pm 3$
coincidence. The average trigger rate is $\sim 70$~Hz, while its efficiency was
estimated from events of different multiplicities produced by Monte Carlo
simulations, resulting in 100\% efficiency for $N_{\mu} \ge 4$.

\section{Data selection and reconstruction}
The data used for the cosmic-ray analysis presented here \cite{Ref:Cosmic2025}
were collected during LHC Run 2 (2015--2018) during periods without circulating
beams in the LHC and with full ALICE magnetic field (0.5 T). The integrated
live time amounts to 62.5 days, during which $>$165M events with at least 1
reconstructed muon in the TPC were accumulated. A multimuon event is defined as
an event with at least 5 muons; the data sample contains 15702 multimuon events,
after the data selections described below.

The TPC tracking algorithm was designed to reconstruct tracks coming out from
the LHC interaction region, working inwards from the outer radius. As a
consequence a cosmic muon crossing the TPC gets reconstructed as two separate
tracks, called {\it up} and {\it down}\/, belonging to the two halves of the TPC
cylinder. A specific algorithm was worked out to match the two track segments
as a single one. Real and Monte Carlo events of different multiplicities were
used to optimize the parameters of this matching algorithm. Moreover, to avoid
possible reconstruction inaccuracies associated with the most inclined tracks,
the zenith angle of all accepted tracks was restricted to $0 < \theta <
50^\circ$\/.

TPC tracks were required to consist of at least 50 clusters (out of a maximum
of 159) and to have a minimum momentum of 0.5~GeV/$c$\/, to eliminate all
possible background from $e^\pm$\/. For multimuon events a parallelism condition
was also applied, which requires the angular difference in space between two
tracks to satisfy $\cos\Delta\Psi > 0.990$\/. Finally, to match up and down
tracks a maximum distance of closest approach of 6~cm in the TPC middle plane
was imposed, this value being chosen to be large enough to maximize the
matching efficiency in high multiplicity events, while keeping combinatorial
background at a minimum. A muon reconstructed with two TPC tracks (up and down)
is called a {\it matched muon}\/; a track satisfying all selection criteria but
the maximum distance is still accepted as muon candidate but flagged as {\it
single-track muon}\/. Most single-track muons are particles crossing the TPC
near the edges where part of their trajectory may fall outside the sensitive
volume.

\section{Muon multiplicity distribution}
\begin{figure}[htb]
     \centering
        \vspace{-0.4cm}
        {\includegraphics[scale=0.480]{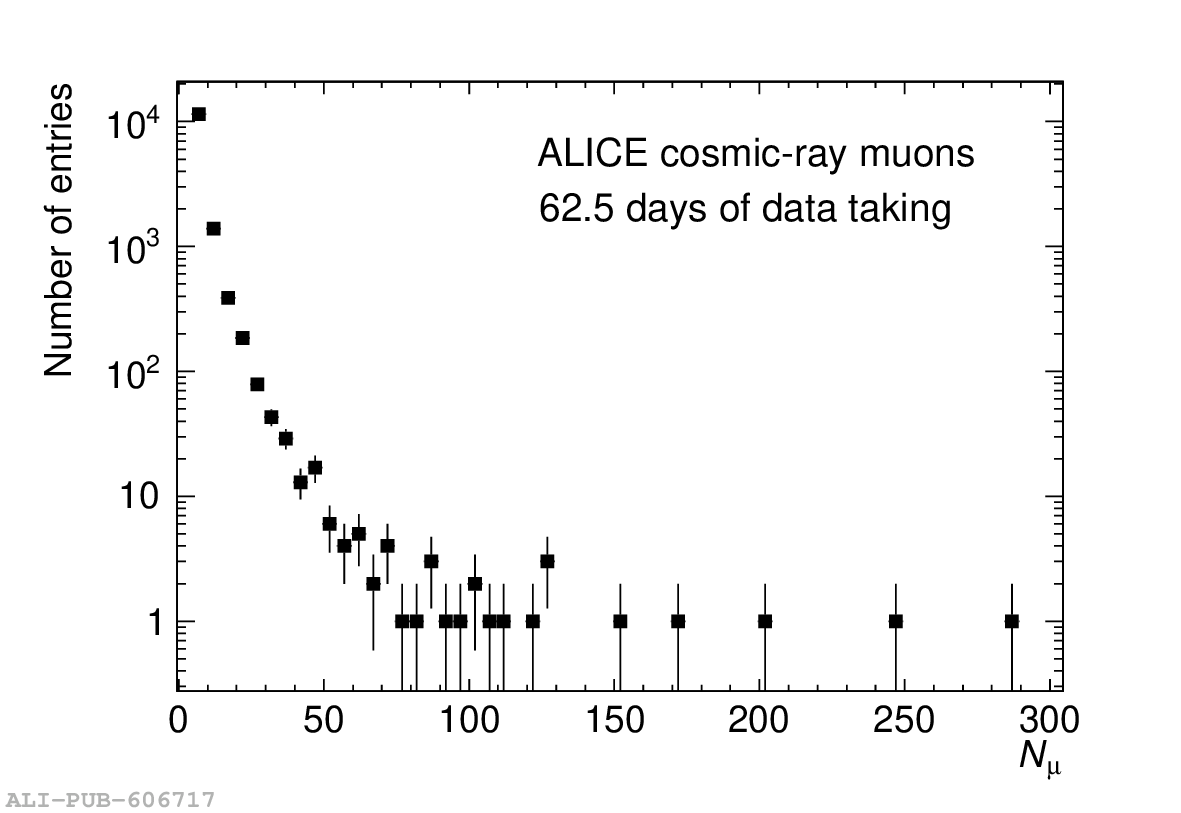}}
        \vspace{-0.4cm}
        \caption{\it Muon multiplicity distribution measured with the ALICE apparatus and obtained for the whole data sample of Run 2. Figure taken from \cite{Ref:Cosmic2025}.}
        \label{Fig:ALICEMMD}
\end{figure}

The muon multiplicity distribution (MMD) obtained from the whole data sample
(Fig.\ \ref{Fig:ALICEMMD}) shows a smooth distribution up to a muon
multiplicity around 50--60, with a well defined decreasing behaviour, and then
13 events with a multiplicity greater than 100. Events with $N_{\mu} > 100$ are
denoted High Muon Multiplicity (HMM) events.

The distribution at low and intermediate multiplicities ($4 < N_{\mu} < 50$),
where the statistical uncertainties are small, is studied by comparing the
measured MMD with simulations. Simulated events equivalent to 62.5 days of
live time were generated using CORSIKA \cite{Ref:CORSIKA} as event generator
and three different hadronic interaction models, QGSJET-II-04 \cite{Ref:QGSJET},
EPOS-LHC \cite{Ref:EPOS} and SIBYLL 2.3d \cite{Ref:SIBYLL}. The ALICE
experimental hall, the surrounding environment and all the detectors are
accurately described using GEANT3 \cite{Ref:GEANT3} for particle propagation.
For each model two samples, pure $p$ (representing a light composition) and
pure Fe (representing an extremely heavy composition), were generated. The
primary cosmic-ray energy was restricted to the interval $10^{14} < E <
10^{18}$~eV following the usual power law energy spectrum $E^{-\gamma}$\/, with a
spectral index $\gamma = 2.70 \pm 0.03$ below the knee ($E_k
= 3\times10^{15}$~eV) and $\gamma = 3.03 \pm 0.03$ above. The total all-particle
absolute flux was extracted from \cite{Ref:AllFlux}. For each shower the core
was randomly scattered at surface level over an area of 205$\times$205~m$^2$
centered above ALICE.

\begin{figure}[htb]
     \centering
        \vspace{-0.2cm}
        {\includegraphics[scale=0.500]{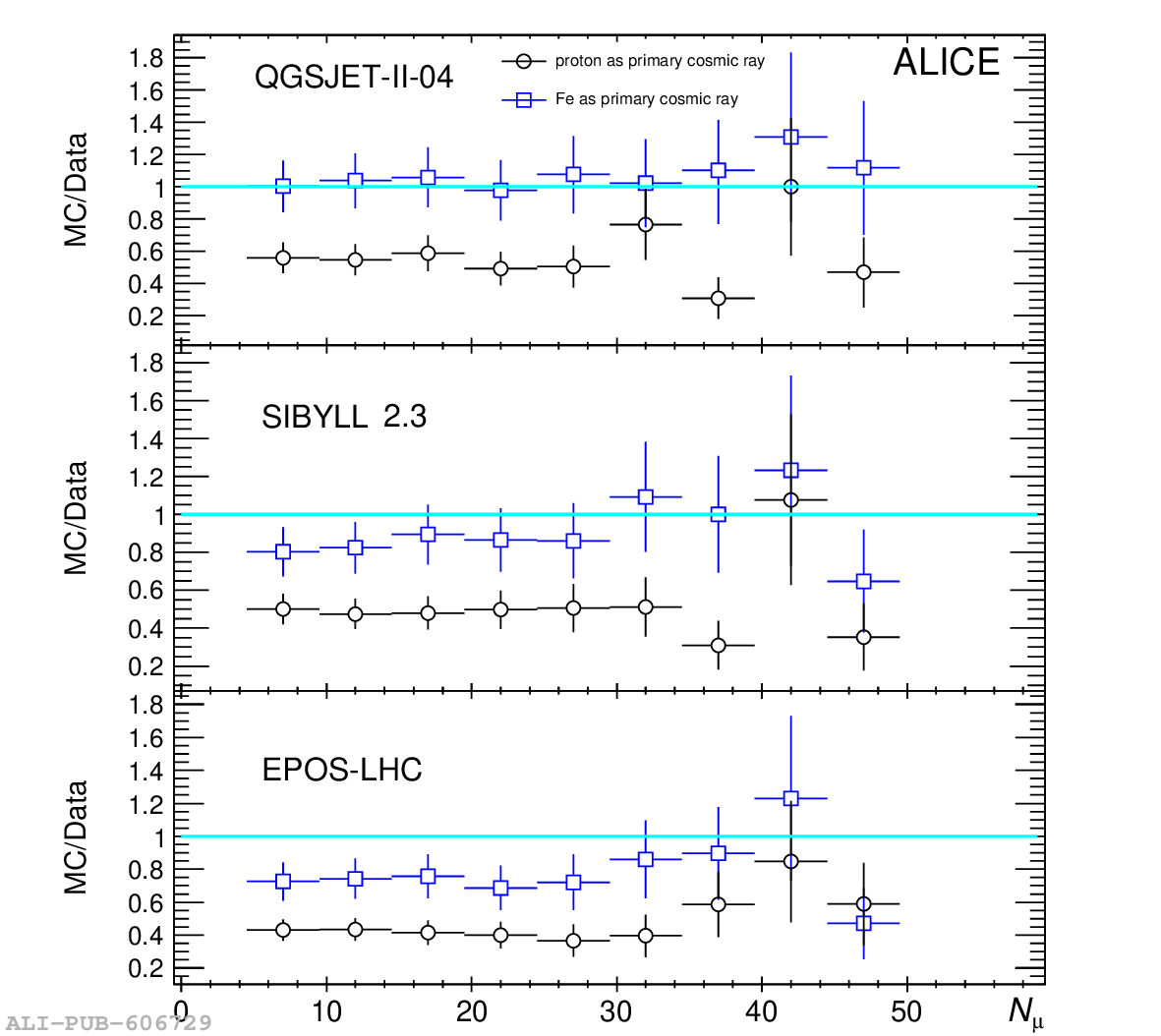}}
        \vspace{-0.2cm}
        \caption{\it Ratio between the number of events obtained with the Monte Carlo simulation with respect to the data for the muon multiplicity distributions. Figure taken from \cite{Ref:Cosmic2025}.}
        \label{Fig:MCData}
\end{figure}

The ratio between the MMD given by the simulations with respect to the data
(MC/Data) is shown in Fig. \ref{Fig:MCData} for the three models. The open
circles represent a pure proton primary composition and the open squares a pure
iron one. Among the three models only QGSJET-II-04 is able to describe the
measured MMD within the whole investigated muon multiplicity interval assuming
a heavy composition of the primary cosmic rays. A study of the energies
involved suggests that the lower multiplicities are due to average energies of
about $4\cdot 10^{15}$~eV increasing up to $> 6\cdot 10^{16}$~eV for
multiplicities greater than 50. On the other hand, neither SIBYLL 2.3d nor
EPOS-LHC can describe the distribution even assuming a heavy composition, both
showing a muon deficit.

\section{Rate of high muon multiplicity events}
In 62.5 live days 13 HMM ($N_{\mu} > 100$) events were recorded, corresponding to
a rate of $(2.4 \pm 0.7)\times10^{-6}$~Hz (i.e.\ 1 HMM event every ($4.8 \pm
1.4$) days). The highest multiplicity cosmic-muon event reconstructed in the
TPC was found to contain 287 muons. To estimate the rate of these events, while
limiting the fluctuations in the number of HMM simulated events, a live time
equivalent of 1 year was simulated. The full simulation was restricted to
primaries with $10^{16} < E < 10^{18}$~eV, since only primaries within this
energy range contribute to these events. To further reduce the statistical
fluctuations, four additional simulations were performed, reusing the same EAS
sample and randomly varying the shower core in the 205$\times$205~m$^2$ area.
Given that the TPC acceptance is some 3000 times smaller, this ensures that the
samples are statistically independent. By averaging the 5 samples the number of
HMM events in 1 year is estimated while reducing the statistical fluctuations.

\begin{table}
  \centering
  \resizebox{\textwidth}{!}{%
  \begin{tabular}[b]{l|cc|cc|cc|cc| p{0.1cm}} 
  \hline
  {} &    &    & \multicolumn{2}{c|}{CORSIKA 7.7400} & \multicolumn{2}{c|} {CORSIKA 7.5600} & \multicolumn{2}{c|} {CORSIKA 7.7400} \\
  HMM events &  Data & Data & \multicolumn{2}{c|}{QGSJET-II-04} & \multicolumn{2}{c|}{EPOS-LHC} & \multicolumn{2}{c|}{SIBYLL 2.3 } \\
  {} & Run 2 & Run 1 & proton & iron & proton & iron & proton & iron \\
  \hline
  \hline
  Period [days per event]            & 4.8 & 6.2 & 10.9 & 5.8 & 15.0 & 9.2  & 13.6  & 6.2  \\
  Rate [$\times 10^{-6}$ Hz]         & 2.4 & 1.9 & 1.1  & 2.0 & 0.8  & 1.3  & 0.9  & 1.9  \\
  Statistical uncertainty            & $28\%$ & $45\%$ & $4\%$ & $5\%$ & $10\%$ & $10\%$ & $10\%$  & $8\%$ \\  
  Uncertainty (syst$\bigoplus$stat)  & $30\%$ & $49\%$  & $23\%$ & $22\%$ & $11\%$ & $16\%$ & $19\%$  & $19\%$ \\
  \hline
  \end{tabular}
  }
  \vspace{2ex}
  \caption{Comparison of the rate of HMM events measured with the data 
  from Run 2 \cite{Ref:Cosmic2025} and Run 1 \cite{Ref:Cosmic2016} and the rate
  calculated with MC simulations of proton and iron samples with the three
  hadronic models. The statistical and total uncertainties are given for
  simulations and data.} 
  \label{Tab:HMMEvents}
\end{table}

\begin{figure}[htb]
     \centering
        \vspace{-0.4cm}
        {\includegraphics[scale=0.570]{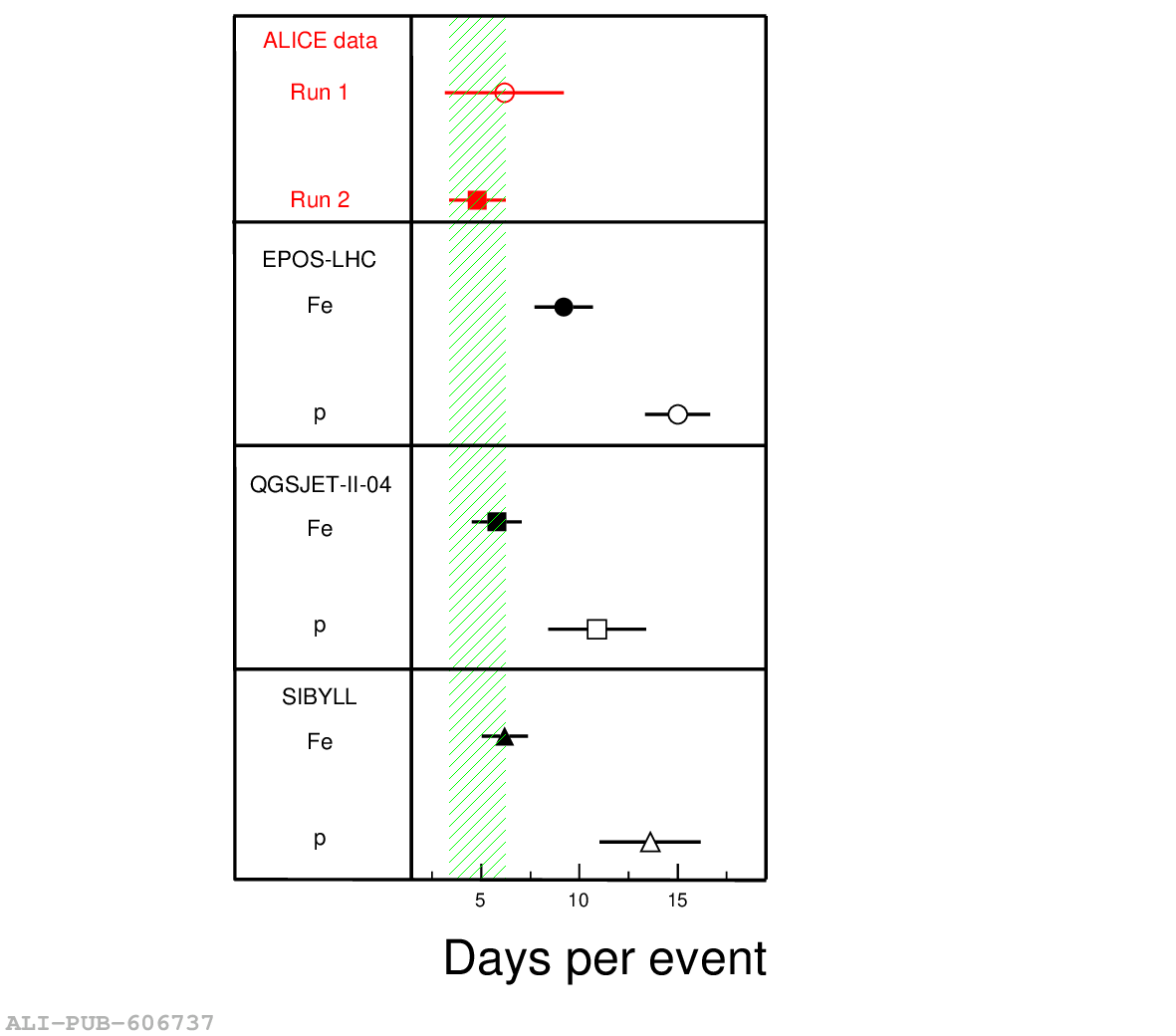}}
        \vspace{-0.4cm}
        \caption{\it  Rate of HMM events (days to yield 1 event) for the data taken in Run 1 and Run 2 compared with the rates obtained with MC simulations with proton and iron samples for the three hadronic interaction models used. The green shaded band is the value of the rate for Run 2 with limits given by 1 standard deviation. Figure taken from \cite{Ref:Cosmic2025}.}
        \label{Fig:HMMSummary}
\end{figure}

In Tab.\ \ref{Tab:HMMEvents} \cite{Ref:Cosmic2025}, the results from Monte Carlo
simulations are compared with data. The rate measured with the Run 2 data
sample is compatible within $1\sigma$ with the one measured during Run 1
(2010--2013) \cite{Ref:Cosmic2016} and has a significantly smaller uncertainty.
For all the three models, the predicted rate for the pure proton composition is
below the data. The rate predicted by QGSJET-II-04 with an iron composition is
the closest to the data and is compatible with the measured rate within
uncertainties. The rate obtained with SIBYLL 2.3d, although lower, is still
compatible with the data while the EPOS-LHC is outside of $2\sigma$. The data
suggest that a composition of cosmic rays that are predominantly heavy elements
is required to explain the measured rate of high multiplicity events. The
estimated average energy of HMM events is about $10^{17}$~eV. These results are
also shown in Fig.~\ref{Fig:HMMSummary} \cite{Ref:Cosmic2025}.

\section{Conclusions}
In 2015--2018, the ALICE Experiment collected 62.5 live days of cosmic-ray data.
The MMD distribution at low and intermediate multiplicities is well reproduced
by Monte Carlo simulations using CORSIKA with the QGSJET-II-04 hadronic model
and assuming a heavy (pure Fe) primary composition, while EPOS-LHC and SIBYLL
2.3d are not able to reproduce the observed distribution. In the same period 13
HMM events were recorded. The observed rate is consistent with the predictions
of the most recent CORSIKA and QGSJET-II-04 models and compatible with SIBYLL
2.3d predictions, while EPOS-LHC is outside of $2\sigma$.


\begin{thebibliography}{99}
\bibitem{Ref:ALEPHRes} V.~Avati {\it et al.} (ALEPH Coll.), Astrop.\ Phys.\
\textbf{19}, 513 (2003).
\bibitem{Ref:DELPHIRes} J.~Abdallah {\it et al.} (DELPHI Coll.), Astrop.\
Phys.\ \textbf{28}, 273 (2007).
\bibitem{Ref:AliceJINST} K.~Aamodt {\it et al.} (ALICE Coll.), JINST
\textbf{3}, S08002 (2008)
\bibitem{Ref:EnerThr} B.Alessandro {\it et al.} (ALICE Coll.), J.\ Phys.\ G:
Nucl.\ Part.\ Phys.\ \textbf{32}, 1295 (2006)
\bibitem{Ref:Cosmic2025} S.\ Acharya {\it et al.} (ALICE Coll.), JCAP {\bf 04},
009 (2025)
\bibitem{Ref:CORSIKA} D.\ Heck {\it et al.} FZKA-6019 (1998)
\bibitem{Ref:QGSJET} S.\ Ostapchenko, Phys.\ Rev.\ \textbf{D83}, 014018 (2011)
\bibitem{Ref:EPOS} T.\ Pierog {\it et al.}, Phys.\ Rev.\ \textbf{C92}, 034906
(2015)
\bibitem{Ref:SIBYLL} E.-J.\ Ahn {\it et al.}, Phys.\ Rev.\ \textbf{D80}, 094003
(2009)
\bibitem{Ref:GEANT3} R.\ Brun {\it et al.}, CERN-DD-EE-84-1 (9, 1987)
\bibitem{Ref:AllFlux} J.\ R.\ H\"orandel, Astrop.\ Phys.\ \textbf{19} 193
(2003)
\bibitem{Ref:Cosmic2016} J.\ Adam {\it et al.} (ALICE Coll.), JCAP {\bf 01},
032 (2016)

\end{thebibliography}
\end{document}